\documentclass[12pt]{article}
\usepackage{graphicx}
\begin{document}

{\bf Scaling Properties, Fractals, and the}

{\bf Renormalisation Group Approach to Percolation}

\bigskip

D. Stauffer

\bigskip
Institute for Theoretical Physics, Cologne University, 

D-50923 K\"oln, Euroland

\bigskip
{\bf \large Article Outline}
\bigskip

Glossary

I. Definition and Introduction

II. Methods

III. Quantities and Exponents

IV. Fractal Dimension; Incipient Infinite Cluster

V. Simple Renormalisation Group

VI. Future Directions

\bigskip
{\bf \large Glossary}
\bigskip

{\bf Cluster}

Clusters are sets of occupied neighbouring sites.

\medskip
{\bf Critical exponents}

At a critical point or second-order phase transition, many quantities diverge
or vanish with a power law of the distance from this critical point; the 
critical exponent is the exponent for this power law.

\medskip
{\bf Fractals}

Fractals have a mass varying with some power of their linear dimension. The
exponent of this power law is called the fractal dimension and is smaller
than the dimension of the space.

\medskip
{\bf Percolation} 

Each site of a large lattice is randomly occupied or empty.

\medskip
{\bf Renormalization}

A cell of several sites, atoms, or spins is approximated by one single site
etc. At the critical point, these supersites behave like the original sites,
and the critical point thus is a fixed point of the renormalisation.

\bigskip
\section{Definition and Introduction}

Paul Flory, who later got the Chemistry Nobel prize, published in 1941 the first
percolation theory \cite{books}, to describe the vulcanisation of rubber 
\cite{flory}. Others later applied and generalised it, in particular by dealing 
with percolation theory on lattices and by studying it with computers. Most of 
the theory presented here was known around 1980, though in the case of 
computer simulation with less accuracy than today. But on the questions of 
universality, of critical spanning probability and of the uniqueness of 
infinite clusters, the 1990's have shown some of our earlier opinions to be 
wrong. And even today it is questioned by some that the critical exponents of
percolation theory can be applied to real polymer gelation, the application
which Flory had in mind two-thirds of a century ago. 
  
On a large lattice we assume that each site independently and randomly
is occupied with probability $p$ and empty with probability $1-p$. Depending on 
applications, also other words can be used instead of occupied and empty, e.g. 
Republican and Democrat for the majority party in an electoral district of the
USA. A {\it cluster} is now defined as set of occupied neighbouring sites. 
Percolation theory deals with the number and structure of these clusters,
as a function of their size $s$, i.e. of the number $s$ of occupied sites in 
the cluster. In particular it asks whether an infinite cluster spans from one
side of the lattice to the opposite side. 
Alternatively, and more naturally if one wants to describe chemical reactions
for rubber vulcanisation, this site percolation can be replaced by bond 
percolation, where every site is occupied but the link between neighbouring 
sites is either present with probability $p$ or absent with probability $1-p$, 
again independently and randomly for each link. A cluster is now a set of 
neighbouring sites connected by links, and the size $s$ of the cluster can be
counted as the number of links, or as the number of sites, in that cluster.
Because of this ambiguity we discuss here mainly site percolation; bond 
percolation is similar in the sense that it belongs to the same universality 
class (same critical exponents). One may also combine both choices and study 
site-bond percolation where each site is randomly occupied or empty, and where
each bond between neighbouring occupied sites is randomly present or absent.

Neither temperature nor quantum effects enter this standard percolation model,
which is purely geometrical probability theory. However, to understand why
percolation works the way it does it is helpful to understand thermal 
phase transitions like the vapour-liquid critical point; and for magnetic 
applications it is useful to know that some spins (atomic magnetic moments)
have only two states, up or down, according to quantum mechanics. We will 
explain these physics aspects later.

For small $p$, most of the occupied sites are isolated $s=1$, coexisting
with only few pairs $s=2$ and triplets $s=3$. For large $p$, most of the 
occupied sites form one ``infinite'' cluster spanning the lattice from left to 
right, with a few small isolated holes in it. Thus there exists one percolation
threshold $p_c$ such that for $p < p_c$ we have no spanning cluster and for
$p > p_c$ we have (at least) one spanning cluster. Inspite of decades of 
research in this seemingly simple problem, no exact solution for $p_c$ is proven
or guessed for site percolation on the square lattice with nearest-neighbour
bonds; only numerically we know it to be about 0.5927462. For site percolation
on the triangular lattice or bond percolation on the square lattice, $p_c=1/2$
exactly. More thresholds are given in Table 1 \cite{books}. They are valid
in the limit of $L \rightarrow \infty$ for lattices with $L^d$ sites in 
$d$ dimensions. For small $L$ instead of a sharp transition at $p_c$ one has 
a rounded changeover: with a very low probability one chain of $L$ occupied 
sites at $p = 1/L^{d-1}$ spans from left to right. In one dimension, a small
chain can easily be spanned if $p$ is close to one, but for $L \rightarrow 
\infty$ the threshold approaches $p_c = 1$ since at smaller $p$ a hole
will appear about every $1/(1-p)$ sites and prevent any cluster to span.

\begin{table}
{\begin{tabular} {lll}  \hline
   $p_c$  & site   &  bond \\ \hline
$d=1$ chain &  1     &   1 \\ \hline
honeycomb & .697043 &   $1-2\sin(\pi/18)$\\
square    & .592746 &    1/2\\
triangular&  1/2   &   $2 \sin(\pi/18)$ \\ \hline
diamond   & .4301   &  .3893\\
SC        & .311608 &  .248813 \\
BCC       & .245691 &  .180287 \\
FCC       & .199236 &  .120163  \\ \hline
$d=4$ hypercubic & .196885 &  .160131 \\
$d=5$ hypercubic & .140797 &  .118172 \\
$d=6$ hypercubic & .109018 &  .094202 \\
$d=7$ hypercubic & .088951 &  .078675 \\ \hline

\end{tabular}}
\caption{Site and bond percolation thresholds for one dimension, three
two-dimensional, four three-dimensional and four hypercubic lattices in higher
dimensions. \cite{books,grassberger}}
\end{table}

\section{Methods}
This section summarises some of the methods employed to find percolation 
properties, first by pencil and paper, and then with the help of computers
for which Fortran programs are published e.g. in \cite{redner,zanjan}. More
details on simulations are reviewed by Ziff in this percolation part of this 
encyclopedia.

\subsection{Mean field limit}

The Bethe lattice or Cayley tree neglects all cyclic links and allows a
solution with paper and pencil. We start from one central site, and let $z$ 
bonds emanate from that. At the end of each bond sits a neighbour. Then from 
each of these neighbours again $z$ bonds emanate, one back to the central site 
and $z-1$ to new sites further outward. They in turn lead again each to $z-1$ 
new sites, and so on. None of the newly added sites agrees with one of the 
already existing sites, and so we can travel along the bonds only outwards or
back, but never in a circle. It is quite plausible that an infinite cluster
of bond percolation is formed if each site leads to at least one more outward
site along an existing bond, that means if $(z-1)p > 1$. This condition also
holds for site percolation. Thus
$$ p_c = 1/(z-1)   \quad . \eqno (1) $$
In this way Flory calculated the threshold and other percolation properties.
Today we call this the ``mean field'' universality class in analogy with 
thermal phase transitions. The critical exponents, to be discussed below,
are integers or simple fractions. To this universality class belong also 
the Erd\"os-R\'enyi random graphs, where we connect in an assembly of $N$ 
points each pair with a low probability $\propto 1/N$. And the same universality
class is reached if we let the dimension $d$ of the hypercubic lattice go 
to infinity (or at least take it above 6). A disadvantage of the Bethe lattice 
is its lack of realism: If the length of the bonds is constant, then the 
exponential increase of the number of sites and bonds with increasing radius 
leads to an infinite density.

\subsection{Small clusters}

The probability of a site to be an isolated $s=1$ cluster on the square
lattice is $n_1 =  p(1-p)^4$ since the site must be occupied and all its four 
neighbours be empty. The formula for pairs is $n_2 = 2p^2(1-p)^6$ since the 
pair can be oriented horizontally or vertically, resulting in the factor 2.
Similar, only more difficult, is the evaluation of $n_s$ with a maximum $s$ 
usually 10 to 20; the general formula is
$$ n_s = \sum_t g_{st} \, p^s \, (1-p)^t \eqno(2)$$
where the perimeter $t$ is the number of empty neighbours and $g_{st}$ is the
number of configurations (or lattice animals, or polyominoes) of size $s$ and 
perimeter $t$. The King's College group in London published these results 
decades ago. With techniques borrowed from series expansions near thermal 
critical phenomena, these polynomials allow to estimate not only $p_c$ but also 
many other quantities (see below) diverging or vanishing near $p_c$. 
 
\subsection{Leath cluster growth}

In the cluster growth method of Leath (1976) one starts with one occupied site
in the centre of the lattice. Then a cluster is grown by letting each empty 
neighbour of an already occupied cluster site decide once and for all, whether
is is occupied or empty. One needs to keep and to update a perimeter list of 
empty neighbours. If that list becomes empty, the cluster growth is finished, 
and no boundary effects of the lattice influence this cluster. If, on the other
hand, the cluster reaches the lattice boundary, one has to stop the simulation
and can regard this cluster as spanning (from the centre to one of the sides). 
Repeating many times this growth simulation one can estimate $p_c$ as well as 
the cluster numbers. More precisely, the cluster statistics obtained in this 
way is not $n_s$ but $n_ss$ since the original centre site belongs with higher
probability to a larger than to a smaller cluster.

\subsection{Hoshen-Kopelman labelling}

To go regularly through a large lattice, which may even be an experimentally 
observed structure to be analysed by computer, one could number consecutively
each seemingly new cluster, and if no clusters merge later then one has a 
clear classification: All sites belonging to the first cluster have label 1,
all sites of the second cluster have label 2, etc. Unfortunately, this does 
not work. In the later analysis it may turn out that two clusters which
at first seemed separate actually merge and form one cluster:

\begin{verbatim}
      * *     * *        1 1     2 2
        *   * * *          1   3 ? 2
      * * * * *          4 ? ? x ?
\end{verbatim}
Already in the simple structure shown on the left we have several such label 
conflicts. The labels to the right come from going though the lattice like a 
typewriter, from left to right, and after each line to the lower line. When
we come to the right neighbour of the 3 we see that 3 is really part of the
cluster with label 2. And at the right neighbour of 4 we see that 4 belongs 
to cluster 1. The stupid method is to go back and to relabel all 3 into 2, 
and all 4 into 1. If then we come to the site marked with x we see that the 
whole structure is really one single cluster, and thus all labels 2 have to 
be relabeled into a 1. This is inefficient for large lattices. Instead, 
Hoshen and Kopelman (1976) gave each site label $m = 1,2,3, \dots$ another 
index $n(m)$. This label $n(m)$ of labels equals its argument, $n(m) = m$, if 
it is still a good ``root label'', and it equals  another number $k$ is the
cluster with initial label $m$ later turned out to be part of an earlier 
cluster $k$. By iterating the command {\tt m = n(m)} until finally the new 
$m$ equals $n(m)$ one finds this root label. For the above we make the following
assignments and re-assignments to $n$: $n(1) = 1, \; n(2) = 2,\; n(3) = 3,\; 
n(3) = 2,\; n(4) = 4, \; n(4)=1, \; n(2) = 1.$ Clusters are now characterised
by the same root label for all their labels. 

An advantage if this method is that only one line of the square lattice, or 
one hyperplane of the $d$-dimensional lattice, needs to be stored at any time, 
besides the array $n(m)$. And that array can also be reduced in size by regular 
recycling no longer used labels $n$, just as beer bottles can be recycled. 
Lattices with more than $10^{13}$ sites were simulated, using parallel 
computers. However, understanding the details of the algorithms and finding 
errors in them can be very frustrating.
 
Sometimes one wants to determine the cluster numbers for numerous different $p$ 
from 0 to 1. Instead of starting a new analysis for each different $p$ one may 
also fill the lattice with new sites, and make the proper labelling of labels 
whenever a new site was added \cite{newziff}. Similarly, one can determine 
the properties of various lattice sizes $L$ by letting $L$ grow one by one and 
relabeling the cluster after each growth step \cite{tigg}. Unfortunately, these 
two methods came long after most of the  percolation properties were already 
studied quite well by standard Hoshen-Kopelman analysis.

\subsection{Relation to Ising  and Potts models}

The relation between percolation and thermal physics was useful for both sides:
Scaling theories for percolation could follow scaling theories for thermal 
physics from ten years earlier, and computer simulations for thermal physics 
could use the Leath and Hoshen-Kopelman algorithms of cluster analysis, leading
to the
Wolff and Swendsen-Wang methods, respectively, a decade later. A mathematical
foundation is given by the Kasteleyn-Fortuin theorem \cite{fortuin} for the 
partition function $Z$ of the $Q$-state Potts model at temperature $T$:
$$ Z(Q)= < Q^N>  \eqno(3)$$
where $N$ is the total number $\sum_s n_s$ of clusters for bond percolation at 
probability $1 = \exp(-2J/k_BT)$, $< \dots>$ indicates an average over the 
configurations at this probability, $k_B$ is Boltzmann's constant and $2J$ is 
the energy needed to break a bond between neighbouring spins. 

$Q$ values of 3 
and larger are interesting since for increasing $Q$ a second-order phase 
transition with a continuous order parameter changes into a first-order phase
transition with a jumping order parameter, when $T$ increases.
The special case $Q=2$ is the spin 1/2 Ising model (the model is pronounced 
EEsing, not EYEsing since Ernst Ising was born in Cologne, Germany, and became 
US citizen Ernest Ising only 
after publishing his theory in 1925 and surviving a Nazi concentration camp.) 
The limit $Q \rightarrow 0$ corresponds to some tree structures (no cyclic 
links, as in Flory's percolation theory, \cite{sokal}). Percolation, on the 
other hand, correponds to the limit $Q \rightarrow 1$, in the following way: The
``free energy'' in units of $k_BT$ is in this limit $\ln Z = \ln < \exp(N \ln 
Q)> \simeq \ln < \exp[(Q-1)N]> \simeq \ln < 1 + (Q-1)N> \simeq (Q-1)N$. Thus for
$Q$ near unity this thermal free energy, divided by $Q-1$, is the 
number of percolation clusters. 

In this way thermal physics and percolation are related, and the cluster 
numbers $N$ correspond to a free energy. In thermal physics, the negative 
derivative of the free energy with respect a conjugate field gives the order
parameter (e.g magnetic field and magnetisation), and the field derivative of 
the order parameter is called the susceptibility. For liquid-gas equilibria,
the order parameter is the volume (or the density), the field is the pressure
(or chemical potential), and the analog of the susceptibility is the 
compressibility. This result Eq.(3), not its derivation, we should keep in mind 
if we now look at the percolation quantities of interest.

Formally we may define for percolation a free energy $F$ as a generating
function of a ghost field $h$:
$$ F(h) = \sum_s n_s \exp(-hs) \quad. \eqno(4)$$
Then its first $h$-derivative is $-\sum_s n_ss$, and the second one is
$\sum_s n_ss^2$, sums which appear below in the percolation probability 
$P_\infty$ (the order parameter) and the mean cluster size 
$S = \sum_s n_ss^2/\sum_s n_ss$ (the susceptibility).

\section{Quantities and Exponents}

The basic quantity is $n_s$, the number (per site) of clusters containing 
$s$ sites each, and often is an average over several realizations for the 
same occupation probability $p$ in the same lattice. Several moments 
$$M_k = \sum_s n_ss^k \eqno(5)$$ are used to define other quantities of 
interest; in these sums the infinite (spanning) clusters are omitted. The 
following proportionalities are valid asymptotically in the limit of large 
lattice size $L$ and for $p \rightarrow p_c$:

$$F = M_0 \propto |p-p_c|^{2-\alpha} + \dots \quad ; \eqno (6a)$$ 
$$P_\infty = p - M_1 \propto (p-p_c)^\beta \quad ; \eqno (6b) $$
$$S = M_2/M_1 \propto |p-p_c|^{-\gamma} \quad . \eqno(6c)$$

Here $F$ is the analog of the thermal free energy, where the three dots 
represent analytic background terms whose derivatives are all finite. 
Since every occupied site must belong either to a finite or to an infinite 
cluster, $P_\infty = p - \sum_s n_ss$ is the fraction of sites belonging to the
infinite cluster and gives the probability that from a randomly selected site
we can walk to a lattice boundary along a path of occupied sites. It is thus 
called the percolation probability but needs to be distinguished from the 
probability $p$ that a single site is occupied and from the probability $R$, 
with $R(p < p_c) = 0, \; R(p > p_c) = 1$, that there is a spanning cluster in 
the lattice. 

The quantity $S$ is usually called the mean cluster size, and we follow this 
tradition even though it is very bad. There are many ways to define a mean size,
and polymer chemists have the much more precise notation of a number average
$M_1/M_0$, a weight average $M_2/M_1$ and a $z$ average $M_3/M_2$ for the
cluster size (= degree of polymerisation). Physicists arbitrarily call the 
weight-averaged $s$ the mean cluster size $S$. Numerically, the exponent 
$\gamma$ is determined more easily from the ``susceptibility'' 
$\chi = M_2 \propto |p-p_c|^{-\gamma}$, since the denominator $M_1$ in Eq.(6b)
approaches very slowly its asymptotic limit of 1.  

The radius of a cluster $R_s$ can be defined as the rms distance $r_i, \; 
i=1,2, \dots, s$ of cluster sites from the centre of mass $r_c$ of the cluster
(radius of gyration):

$$R_s^2 = <\sum_i (r_i-r_c)^2/s> \eqno(6a) $$
where the $<\dots>$ are the average over all cluster configurations at 
probability $p$. Then the correlation length $\xi$ is related to the 
$z$-average cluster radius through
$$\xi^2=\sum_s R_s^2 n_s s^2/\sum_s n_s s^2 \propto |p-p_c|^{-\nu} \eqno (6b)$$
with another critical exponent $\nu$. 

Finally, right at $p=p_c$, the cluster numbers decay as
$$ n_s \propto 1/s^{2 + 1/\delta}  \eqno (7)$$  
where $\delta$ must be positive to allow a finite density $\sum_s n_ss = p$.

These five critical exponents are not independent of each other but are related
in $d$ dimensions through the scaling laws:

$$2-\alpha = \gamma + 2 \beta = (\delta+1)\beta = d\nu \eqno (8a)$$
as known from thermal phase transitions; the last equation involving $d$ is not
valid in mean field theory (large $d$) but only for $d \le 6$. Table 2
gives the numerical estimates of the exponents in three dimensions as well as 
their mean field values for $d \ge 6$ and their exact two-dimensional results
\cite{nienhuis,smirnov}. Thus, for six and less dimensions, if you know 
two exponents you know them all; thus far.
 
\begin{table}
{\begin{tabular} {llll}  \hline
$d$  & $\beta$   &  $\gamma$ & $\nu$ \\ \hline
2    &  5/36     &    43/18  & 4/3   \\
3    &  0.41     &    1.796  & 0.88  \\
$\ge 6$ & 1      &     1     & 1/2   \\

\end{tabular}}
\caption{Critical exponents for percolation clusters. The mean field values are 
valid for six and more dimensions and also apply to Flory's Bethe approximation
and to Erd\"os-R\'enyi random graphs. The exponents $\alpha, \, \delta, \, 
\sigma, \, \tau$ can be derived from the scaling laws, Eq.(8).}
\end{table}

These scaling laws (8a) can be derived by assuming 
$$n_s = s^{-\tau} f[(p-p_c)s^\sigma] \quad (\tau=2+1/\delta, \; 
1/\sigma=\beta\delta) \eqno(8b) $$
which was first postulated for the thermal Ising model, and then successfully
applied to percolation. Here $f$ is a suitable scaling function, which only 
in the mean-field limit approaches a Gaussian.

For both thermal critical phenomena and percolation, ``universality'' asserts
that these critical exponents are independent of many details and (for the 
Potts model) depend only on the dimensionality $d$ and the number $Q$ of 
possible spin states. Since percolation corresponds to $Q \rightarrow 1$ this
means that the exponents depend only on $d$. There are exceptions from this 
universality for thermal phase transitions, but for random percolation thus far
it worked. However, the numerical value of the percolation threshold $p_c$ is 
not a critical exponent, depends on the lattice structure, and is different 
for site and bond percolation. 

This universality is one of the reasons why the investigation of exponents is
important: They allow to classify models and materials. Similarly, in biology 
we have many birds of different colours, and many types of domestic animals.
Biology became a systematic science only when it was found that all mammals 
share certain properties, which birds no not have. Thus there is the
universality class of mammals. 
 
(The proportionality factors in Eq.(6) are not universal, but some of their 
combinations are; for example, the ratio of the proportionality factors for
$S$ above to below $p_c$ is universal. In some sense also the probability 
$R(p=p_c)$ of a lattice to contain one spanning cluster at the threshold is
universal: same for bond and site percolation; however, that probability 
depends on the boundary conditions and the shape of the sample and thus is far
less universal that the mentioned ratio for $S$.)

Unfortunately, there is another exponent which does not follow 
from the cluster numbers and radii and for which no scaling law is accepted
which relates it to the other exponents above. This refers to the electrical
conductivity $$\Sigma \propto (p-p_c)^\mu \eqno (9) $$
when each occupied site (or bond) conducts electrical current and each
empty site (or deleted bond) is an insulator. The numerical values are 1.30,
2.0 and 3 in two, three and at least six dimensions. If bonds are related 
by elastic springs with bending forces, the elastic exponent may be
$\mu + 2\nu$ if entropy effects are negligible, or $2-\alpha$ if entropy effects
are dominant. Moreover, $\mu$ is less universal: the above lattice values do not
hold on a continuum (conducting spheres which may overlap). Similarly, the 
kinetics of the Ising model determine a critical exponent which may differ
in different variants of the kinetics and may not be related to the static
Ising exponents like $\beta$ and $\gamma$. 

\section{Fractal dimension; incipient infinite cluster}
\subsection{Fractal dimension $D$}

Typical objects of geometry classes in school are one-dimensional lines, 
two-dimensional squares or circles, and three-dimensional cubes or spheres.
They have a length (radius) $L$ and a mass (volume for unit density) $M$ with
$M \propto L^d$ for $d$ dimensions. In reality, mother nature produces more more
complex objects, like trees, where the mass varies with a power of the
tree height below 3:

$$M \propto L^D \quad (D < d, \; L \rightarrow \infty). \eqno(10a)$$
$D$ is the fractal dimension, and such objects are called fractals, particularly
if they also are self-similar in that a small twig looks like a big branch, etc.
Finite-size scaling theory then relates $D$ of the largest
(spanning?) cluster at $p = p_c$ to the above percolation exponents through
$$D = d-\beta/\nu=(\gamma+\beta)/\nu=1/(\sigma \nu)=d/(1+1/\delta) \quad  
\eqno (10b)$$
for $d \le 6$. Thus the critical cluster is about 1.9-dimensional in two and 
2.5-dimensional in three dimensions, while in the mean field regime for
$d \ge 6$ we have $D = 4$. 
  
Why is this so? Any quantity $X$ which is supposed to vary near $p=p_c$ as
$|p-p_c|^x$ does so only for infinitely large systems. For a finite lattice 
size $L$, the transition is rounded, and neither $X$ nor any of its 
$p$-derivatives diverges or becomes exactly zero. In particular, the typical
cluster radius or correlation length $\xi \propto |p-p_c|^{-\nu}$  cannot 
become infinite but becomes of order $L$. Then the relation $X \propto 
\xi^{-x/\nu}$ is replaced by 
$$ X(p=p_c) \propto L^{-x/\nu} \eqno (11a)$$
at the threshold, and 
$$ X(p \simeq p_c) = L^{-x/\nu} g[(p-p_c)L^{1/\nu}] \eqno (11b)$$
near the threshold, with a suitable scaling function $g$. In particular, 
the fraction $P_\infty$ of sites belonging to the largest cluster at $p = p_c$ 
vanishes as $L^{-\beta/\nu}$, and the total number $M$ of sites in this 
cluster as

$$ M \propto L^{d-\beta/\nu} \quad {\rm or} \quad D = d - \beta/\nu \eqno (11c)$$ 
as asserted in Eq.(10b).

Fig.1 shows the second moment $\chi = M_2 = \sum_s n_ss^2$ in small (curve) and 
large (+) simple cubic lattices, differing only for $p \simeq p_c$.

In a finite lattice, the probability $R(p)$ of a spanning cluster to exist
goes from nearly zero to nearly unity in a $p$-interval proportional to 
$1/L^{1/\nu}$, according to Eq.(11a) with $x = 0$. The derivative d$R$/d$p$ 
is the probability that spanning first occurred at probability $p$. It is 
plausible that this probability, peaked around $p_c$, is a Gaussian, i.e.
a normal distribution. Unfortunately, the Evil Empire, also known as the
Departments of Chemical Engineering, destroyed \cite{ziff94} this beautiful 
idea: Since for $p \simeq p_c$ and $\xi \sim L$ every part of the lattice is
correlated with the rest of the lattice, the central limit theorem does not 
hold.
 
(If for $p \ll p_c$ we let the cluster size $s$ go to infinity, which requires 
a special algorithm, we get into the universality class of lattice animals,
Sec.(2.2). Most simply, in the limit $p \rightarrow 0$, Eq.(2) simplifies
to $n_s/p^s = g_{st}$, that means we look at the distribution of configurations 
with $s$ sites and perimeter $t$, where all configurations of a given $s$ are 
weighted equally, whatever their perimeter $t$ is. An important result for
these animals is that in three dimensions their radius $R_s$ varies as $\sqrt 
s$, i,e. their fractal dimension is exactly 2. In two dimensions, only numerical
estimates exist with $D \simeq 1.56$. It is highly unusual that a problem has
an exact solution in three but not in two dimensions.)
 
\begin{figure}[hbt]
\begin{center}
\includegraphics[angle=-90,scale=0.5]{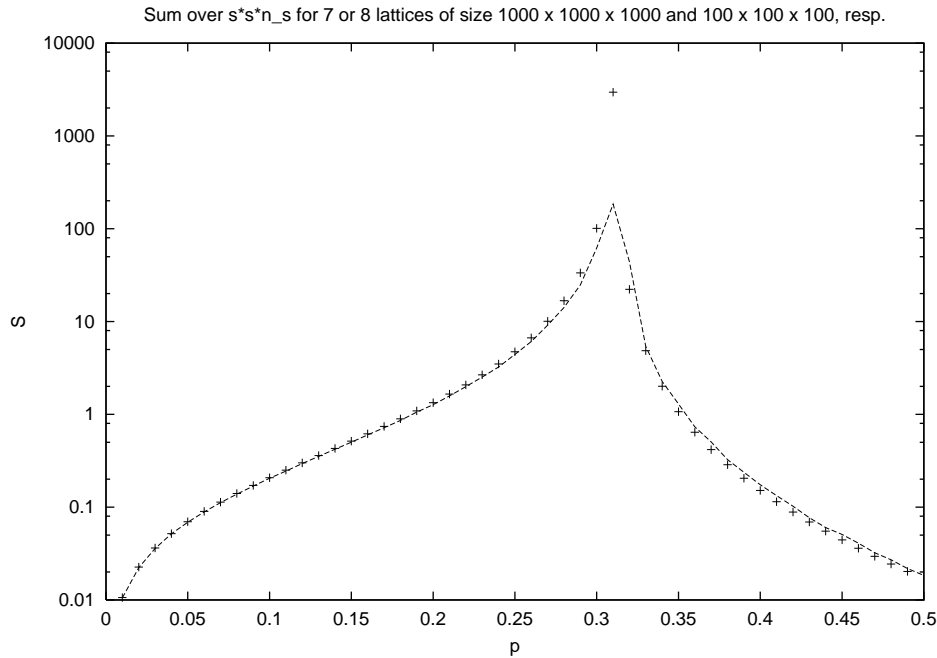}
\end{center}
\caption{``Susceptibility'' $M_2$ in simple-cubic lattice. For the smaller
size the maximum is reduced appreciably.
}
\end{figure}

\begin{figure}[hbt]
\begin{center}
\includegraphics[angle=-90,scale=0.5]{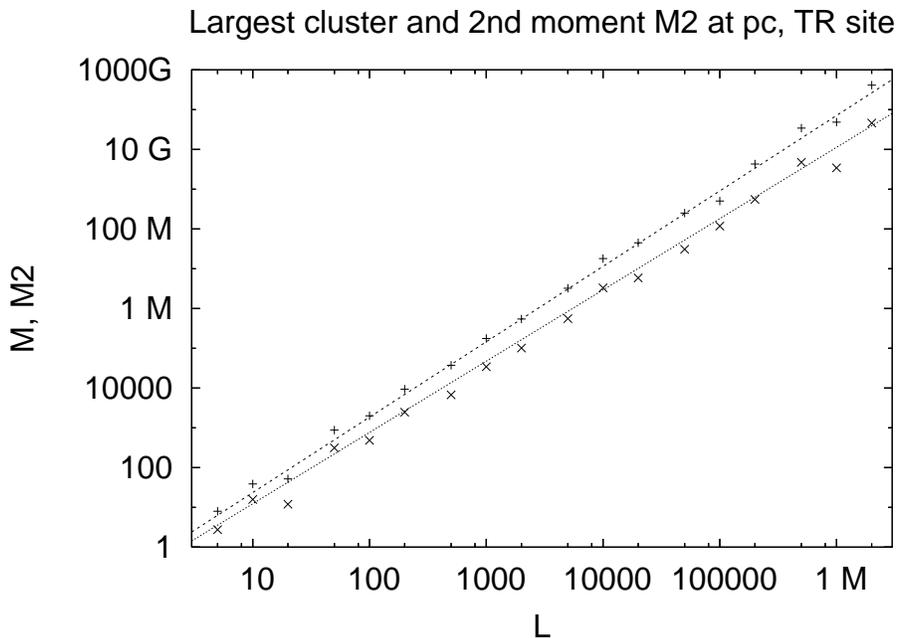}
\end{center}
\caption{Number $M$ of sites in largest cluster (+) and susceptibility $M_2$ (x)
at $p=p_c= 1/2$ for triangular site percolation. The two straight lines have the
exact slopes $D = 91/38$ and $\gamma/\nu = 43/24$ predicted by finite-size 
scaling. The largest lattice took about 36 hours on a workstation with 2 
Gigabytes memory. Tiggemann \cite{tigg} simulated $L = 7 \times 10^6$, 25024, 
1305, 225 for $d = 2$, 3, 4, 5 on a large parallel computer.
}
\end{figure}


\subsection{Incipient infinite cluster}

Right at $p=p_c$ the largest cluster spans the lattice with a pseudo-universal
probability  $0 < R(p_c) <1$, and then has a density $P_\infty$ going to zero
for $L$ going to infinity. It is also called the incipient infinite cluster 
IIC. Most of the IIC consists of dangling ends which carry no current if the
cluster is interpreted as a random resistor network with conductivity $\Sigma$,
see Eq.(9) above. The remaining current carrying ''backbone'' has a fractal 
dimension 
1.643 in two dimensions, 1.7 in three and 2 in at least six dimensions and 
mostly consists of blobs where current flows along several parallel though
connected paths. The few ``articulation'' sites or bonds, the removal of which 
cuts the network into two or more parts, are also called ``red'' since all the 
current flows through them; they have a fractal dimension of only $1/\nu$ = 
0.75, 1.14 and 2 in two, three and $\ge$ six dimensions.

How many infinite clusters do we have? The easy answer is: none below, perhaps
one at and always one above $p_c$ in an infinite network. Indeed, this is 
what was claimed mathematically in the 1980's \cite{newman}: The number of 
infinite clusters is zero, one or infinite. Later mathematics excluded the last
choice of infinitely many clusters, even though in seven dimensions scaling
arguments, confirmed by numerical studies \cite{arcan}, indicated the number of
IIC to go to infinity for increasing $L$ in seven dimensions. Only in 1995
and later Aizenman \cite{aizenman} predicted that in all dimensions one
may have several spanning clusters at $p = p_c$, in agreement with simulations
\cite{shchur}.

Why were the earlier uniqueness theorems irreproducible at $p_c$ and for
very elongated rectangles even above $p_c$ \cite{fisher}? A clear definition
of ``infinite'' is missing in some of the mathematics, although \cite{newman}
defined a cluster as infinite if its cardinality (= number of sites in it) is
infinite for $L \rightarrow \infty$ in a hypercubic lattice of $L^d$ sites.
Clear definitions of infinity are, of course, needed for reliable proofs
\cite{jarai}. Measure theory as applied in some theorems may be based on some 
axioms which are not applicable for a fractal IIC. Very simply, imagine each 
line of an $L \times L$ square lattice to have one randomly selected site 
occupied and all others empty. The set of occupied site then has cardinality
$L$ which is infinite for infinite lattices, but its density becomes zero. Does 
your measure theory agree with
this? More relevant for percolation, even for $p < p_c$ the largest cluster
has a size increasing logarithmically with lattice size and thus can be
described as infinite, invalidating the percolation threshold as the onset of 
infinite clusters. Thus infinite might be defined as increasing with a 
positive power of $L$, i.e. having a positive fractal dimension. Then we have 
infinitely many infinite clusters only at $p=p_c$, though in most cases only the
largest of them is a spanning cluster. Using ``spanning'' as a definition of an 
infinite cluster seems to cause the smallest problems.
 
Thus one should not regard a question as settled if some mathematical theorem 
claims to have answered it. The mathematics may not apply to the same problem 
one is interested in, or (see bootstrap percolation in this encyclopedia)
may apply only for unrealistically large lattices. On the other hand, also 
computer simulations should be relied upon only if confirmed independently.
And in the interpretation of simulation results one should be objective and 
not try to agree with prevailing theories. For example, \cite{arcan} might
already have seen the multiplicity of infinite clusters in five dimensions,
not only in seven, had she not followed her obviously incompetent postdoctoral
mentor. 

(On a more positive side, mathematicians \cite{ganten} solved biased diffusion 
on percolating clusters above $p_c$ only a few years after physicists still 
had controversies about their simulations.)

\section{Simple Renormalisation Group}

Why are scaling laws and finite-size scaling so simple? Why is universality 
valid for the exponents? These question arose for thermal critical phenomena as 
well as for percolation. The main reason is that the correlation length $\xi$ 
goes to infinity at the critical point. Thus all approximations which restrict
the correlation to some finite lengths eventually become wrong, and instead the
scaling ideas become correct. They were explained by Ken Wilson through what he 
called renormalisation group, around 1970, and he got the physics Nobel prize 
for it in 1982. Basically, since correlations extend over long distances, the
single atom or lattice point becomes irrelevant and can be averaged over. 
In politics, we have a similar effect: Many democracies are based on electoral
districts, and the candidate winning most votes within this district represents 
this district in the national parliament. It is the cooperation of many people 
within the electoral district, not the single vote, which is decisive.

Returning to an $L \times L$ lattice, we can divide it into many blocks of 
linear dimension $b$, and treat a block analogously to an electoral district. 
Thus in an Ising model, if the majority of block spins point upward, the whole 
block is represented by a superspin pointing up, analogous to the single 
representative in politics. These block spins then act like the original spins, 
one can put $b \times b$ superspins into one superblock, and have just one 
super-representative following the majority opinion of the representatives 
within the superblock. This process can be continued: at each stage $b \times 
b$ lower representatives are renormalised into a single higher representative. 

Such a renormalisation by majority rule works fine with Ising spins, but 
percolation deals with connections, not with up and down spins. Thus for percolation
a $b \times b$ block is renormalised into an occupied supersite if and only 
if there is a spanning cluster within the block; otherwise the superblock is 
defined empty. In this way, whole blocks are renormalised into singe sites
via connectedness. And the renormalisation is reduced to the standard 
question which was asked already before Wilson's invention: Does a $b \times
b$ lattice have a spanning cluster?  The supersite is thus occupied if and only
if the block spans, which happens with probability $R_b(p)$. If we call the
$p'$ probability of the supersite to be occupied, we thus have

$$p' = R_b(p) \quad.  \eqno(12a) $$
If we are at $p = p_c$, then the renormalisation should not change anything 
drastic since $\xi$ is larger than any $b$; thus if the renormalisation would
be exact we would have 

$$p_c' = R_b(p_c) \quad . \eqno(12b)$$
Practically we determine a fixed point $p = p^*$ such that

$$p^* = R_b(p^*) \quad . \eqno(12c)$$
and then find $p_c$ as the limit of $p^*$ for $b \rightarrow \infty$, which again
is similar what percolation experts did before renormalisation theory.

A particularly simple example is the triangular site percolation problem with
$p_c = 1/2$, if we do not divide the lattice into large $b \times b$ blocks but 
into small triangles of three sites which are nearest neighbours, as shown on 
the left:

\begin{verbatim}
   *           x          x
 *   *       x   x      x   .
\end{verbatim}

The triangle contains a spanning cluster if either all three sites are 
occupied (x, central diagram) or two sites are occupied (x) and one site is
empty (. , right diagram). The first choice appears with probability $p^3$, the
second with probability $p^2(1-p)$. However, this second choice has three
possible orientations since each of the three sites can be the single empty
site. Thus the total probability of the triangle to have a spanning cluster 
is 
$$p' = p^3 + 3 (1-p)p^2 \eqno (13a)$$
with three fixed points $p^*$ where $p' = p$:

$$p^* = 0, \quad p^* = 1/2, \quad p^* = 1 \quad . \eqno(13b)$$
The second of these fixed points is the percolation threshold, while the
first correponds to lattice animals (section 2.2 and end of section 4.1) and 
the third to compact non-fractal clusters. With somewhat more effort one can
derive also a good approximation for $\nu$.

This agreement of the fixed point $p^*$ with the true threshold $p_c = 1/2$
is not valid for other lattices or block choices. Nevertheless there was a
widespread fixed-point consensus that $R_b(p_c) = p_c$ for sufficiently large 
$b$. Regrettably, the Evil Empire \cite{ziff92} again destroyed this beauty and
found $R_b(p_c) = 1/2$ for square site percolation where $p_c \simeq 0.593$. 
In general, $R(p_c)$ is a pseudo-universal quantity depending on boundary 
conditions and sample shape, while $p_c$ for large samples is independent
of these details but is different for site and bond percolation and depends on 
the size of the neighbourhood. Life was much nicer before. Fortunately, if a 
fixed point is determined by Eq.(12c) and the block size goes to infinity,
then the fixed point still approaches $p_c$.

\section{Future Directions}
This review summarised the basic theory, particularly
when it was not yet contained in the earlier books \cite{books}. Applications
were left to the Sahimi book \cite{books}; even for the very first
application \cite{flory} there is not yet a complete consensus that the 
three-dimensional percolation exponents apply to polymer gelation. More recent 
applications are social percolation \cite{weisbuch} for marketing 
by word-of-mouth, and stock market fluctuations due to herding among traders
\cite{cont}. 

Percolation theory, similar to Fortran programming or capitalism, was thought 
to be finished but seems to be alive and kicking. Nevertheless I think the
future is more in its applications. 
 
The manuscript was improved by criticism of A. Aharony.


\begin{thebibliography}{99}
\bibitem{flory}
{\bf Primary literature:}
 
\bigskip
Flory PJ (1941) J. Am. Chem. Soc. 63: 3083

\bibitem{grassberger} Grassberger P (2003) Phys. Rev. E 67: 036101

\bibitem{redner} Redner S (1982) J. Statist. Phys. 29: 309

\bibitem{zanjan} Stauffer D, Jan N (2000) In: {\it Annual Reviews of
Computational Physics}, vol. VIII (Zanjan School), World Scientific,
Singapore.

\bibitem{newziff} Newman MEJ, Ziff RM (2000) Phys. Rev. Lett. 85: 4104

\bibitem{tigg} Tiggemann D (2006) Int. J. Mod. Phys. C 17: 1141 and PhD 
thesis, Cologne University.

\bibitem{sokal} Deng VJ, Garoni TM, Sokal, AD (2007) Phys. Rev. Lett. 98: 
030602

\bibitem{fortuin} Kasteleyn PW, Fortuin CM (1969  J. Phys. Soc. Jpn. Suppl.
S 26: 11

\bibitem{nienhuis} Nienhuis B (1982) J. Phys. A 15: 199

\bibitem{smirnov} Smirnov S, Werner W (2001) Mathem. Res. Lett. 8: 729

\bibitem{ziff94} Ziff RM (1994) Phys. Rev. Lett. 72: 1942

\bibitem{newman} Newman CM, Schulman LS (1981) J.Stat.Phys. 26: 613

\bibitem{arcan} de Arcangelis L (1987) J.Phys. A 20: 3057

\bibitem{aizenman} Aizenman M (1997) Nucl.Phys. (FS) B 485: 551

\bibitem{shchur} Shchur LN, Rostunov T (2002) JETP Lett. 76: 475

\bibitem{fisher} Stauffer D (1999) J. Irreproducible Results 44: 57
                       
\bibitem{jarai} Jarai AA (2003) Ann. Prob. 31: 444

\bibitem{ganten} Berger N, Ganten N, Peres Y (2003) Probab. Theory Relat. 
Fields 126: 221

\bibitem{ziff92} Ziff RM (1992) Phys. Rev. Lett. 69: 2670

\bibitem{weisbuch} Weisbuch G, Solomon S (2002) In: Bornholdt S, Schuster
HG (eds) {\it Handbook of graphs and networks}, Wiley-VCH, Weinheim, p.113

\bibitem{cont} Cont R, Bouchaud J-P (2000) Macroeconomic Dynamics 4: 170 

\bigskip
\bibitem{books} 
{\bf Books and Reviews:}

D. Stauffer, Phys. Repts. 54, 1 (1979); J.W. Essam, Repts.
Progr. Phys. 43, 843 (1980); D. Stauffer and A. Aharony, 
{\it Introduction to Percolation Theory}, Taylor and Francis, London
1994 (revised second edition); M. Sahimi, {\it Applications of Percolation 
Theory}, Taylor and Francis, London 1994; A. Bunde and S. Havlin, {\it Fractals
and Disordered Systems}, Springer, Berlin 1996; G. Grimmett, {\it Percolation},
second edition, Springer, Berlin 1999.

\end{thebibliography}
\end{document}